\documentclass[twocolumn,prl]{revtex4}
\usepackage{float}
\usepackage{makeidx}
\usepackage{graphicx}

\begin{document}

\title{Nonlocal Nondegenerate Optical Parametric Amplifier Based on Genuine Multipartite Entanglement}
\author{Jing Zhang$^{\dagger }$, Changde Xie, Kunchi Peng}
\affiliation{State Key Laboratory of Quantum Optics and Quantum
Optics Devices, Institute of Opto-Electronics, Shanxi University,
Taiyuan 030006, P.R.China}

\begin{abstract}
We propose a new scheme for realizing nonlocal nondegenerate optical
parametric amplifier by means of genuine four-mode entanglement. The
nondegenerate optical parametric amplifier is regarded as a unitary
transformation from two input fields to two outputs. Two inputs and
two outputs are located at the remote stations respectively.
Employing off-line four-mode entanglement, homodyne detection,
classical communication and local displacement, the nonlocal
nondegenerate optical parametric amplifier can be achieved. This
scheme can be implemented with a setup that is experimentally
accessible at present.
\end{abstract}

\maketitle

The nondegenerate optical parametric amplifier (NOPA) can be
described with the following input-output unitary transformation

\begin{eqnarray}
\hat{a}_{out}^{s}
&=&\sqrt{G}\hat{a}_{in}^{s}+\sqrt{G-1}\hat{a}_{in}^{i\dagger},
\label{PIA1}\\
 \hat{a}_{out}^{i}
&=&\sqrt{G-1}\hat{a}_{in}^{s\dagger}+\sqrt{G}\hat{a}_{in}^{i},
\label{PIA2}
\end{eqnarray}
where $\hat{a}_{in}^{s(i)}$ and $\hat{a}_{out}^{s(i)}$ represent the
annihilation bosonic operators of the input and output signal
(idler) fields respectively, and $G$ is the power gain. Usually NOPA
is closely related to phase-insensitive amplification, which only
involves Eq. (\ref{PIA1}). The quantum state $\hat{a}_{in}^{s}$ is
amplified by an amount $\sqrt{G}$, and at the same time an additive
noise $\sqrt{G-1}\hat{a}_{in}^{i\dagger}$ must be attached to
maintain commutation relation
$([\hat{a}_{out}^{s},\hat{a}_{out}^{s\dagger}]=1)$ of the amplifier
output. This ultimate limits imposed by quantum mechanics were
addressed long years ago \cite{one,two,three}. This intrinsic noise,
intimately linked with measurement theory and the no-cloning
theorem, gives rise to many inextricable restrictions on the
manipulations of quantum states. Despite these limitations, the
phase-insensitive amplifier is rich of application, in particular,
in optical communication. Furthermore, NOPA is of particular
importance for quantum information and communication, for example
the generation of two-mode Gaussian entangled state
(Einstein-Podolskyo-Rosen (EPR) entangled state) \cite{four,five},
and quantum cloning \cite{six,seven,eight}.

Numerous apparatuses can be used for accomplishing the
phase-insensitive amplification, for instance, solid state laser
amplifiers \cite{nine}, and the schemes based on the linear optics
and feedforward \cite{ten}. However, the idler output in these
schemes is not always accessible, e.g., in a laser amplifier this
mode is scattered into vibrational modes of atoms. Although NOPA can
be achieved directly by type-II parametric interaction \cite{four,
five}, there is a lot of technic trouble in the implementation
process for coupling efficiently, mediated by a nonlinearity, the
input signal and idler fields into the pump field. We define this
type of NOPA as the local NOPA since the two input and two output
fields are at a same location. The localization limits its
application in quantum information and communication. A recent trend
in quantum information science is that the nonlinear interaction are
efficiently replaced by off-line entanglement, classical
communication, and local operation \cite{eleven, twelve}. The
protocol of substituting the in-line parametric amplifier with
linear optics, feedforward and an off-line EPR source has been
suggested \cite{ten,thirteen1,thirteen}, such that the trouble to
couple the input signal and idler fields with the pump field in a
nonlinear crystal can be overcome. In Ref. \cite{thirteen1} local
NOPA and quantum nondemolition (QND) interaction based on off-line
squeezers, linear optics, and measurements is proposed. Recently,
implementing nonlocal interaction between spatially distant quantum
network nodes becomes very interesting work. Nonlocal QND
interaction can be implemented \cite{fifteen1} by using off-line
four-mode cluster state \cite{fifteen}, which is the genuine
multipartite entanglement. In this paper we show that NOPA can be
achieved nonlocally by using off-line four-mode entanglement,
homodyne detection, classical communication and local displacement
to replace the in-line nonlinear interaction. In the proposed
system, all two inputs and two outputs locate four remote stations
respectively. The used off-line four-mode entanglement is a new
class of genuine multipartite entangled states, which is different
from Greenberger-Horne-Zeilinger (GHZ) and cluster states and can be
generated by combining two beams from two different pairs of EPR
entangled beams on a beam splitter.

First we must prepare a four-mode entangled state and distribute
them to four remote stations. As shown in Fig. 1, two pairs
$(\hat{a}_{EPR1},\hat{a}_{EPR2})$ and
$(\hat{b}_{EPR1},\hat{b}_{EPR2})$ of EPR entangled states are
utilized to generate the genuine four-mode entangled state. The EPR
entangled beams have the very strong correlation property,
such as both their sum-amplitude quadrature variance $\langle \delta (\hat{X}%
_{a(b)_{EPR1}}+\hat{X}_{a(b)_{EPR2}})^2\rangle =2e^{-2r_{1(2)}}$,
and their
difference-phase quadrature variance $\langle \delta (\hat{Y}_{a(b)_{EPR1}}-%
\hat{Y}_{a(b)_{EPR2}})^2\rangle =2e^{-2r_{1(2)}}$, are less than the
quantum noise limit, where $r_1$ and $r_2$ are the squeezing factor,
and, $\hat{X}$ and $\hat{P}$ describe the amplitude and phase
quadrature of optical field with the canonical commutation relation
$[\hat{X},\hat{P}]=2i$. One of the EPR entangled beams
($\hat{a}_{EPR2}$) is combined with one of the other EPR pair
($\hat{b}_{EPR2}$) on a beam splitter $BS_{0}$ with reflectivity
rate $R$ and transmission rate $T=1-R$. The output fields are
expressed by
\begin{eqnarray}
\hat{c}_{t}=\sqrt{1-R}\hat{b}_{EPR2}-\sqrt{R}\hat{a}_{EPR2},\nonumber \\
\hat{c}_{r}=\sqrt{R}\hat{b}_{EPR2}+\sqrt{1-R}\hat{a}_{EPR2}.\label{multi}
\end{eqnarray}
So, the four-mode entangled state for nonlocal NOPA is described
with the annihilation operator $\hat{a}_{1}=\hat{a}_{EPR1},
\hat{a}_{2}=\hat{c}_{t}, \hat{a}_{3}=\hat{c}_{r},
\hat{a}_{4}=\hat{b}_{EPR1}$. The properties of the four-mode
entangled state for nonlocal NOPA will be discussed in the following
by its quantum correlation variances of the amplitude and phase
quadratures (position and momentum). The quantum correlations of the
four-mode entangled state for nonlocal NOPA can be obtained easily
with
$\langle\Delta ^2(X_{a_1}-\sqrt{R}X_{a_2}+\sqrt{1-R}X_{a_3})\rangle<SNL$, $%
\langle\Delta
^2(\sqrt{1-R}X_{a_2}+\sqrt{R}X_{a_3}+X_{a_4})\rangle<SNL$
and $\langle\Delta ^2(P_{a_1}+\sqrt{R}P_{a_2}-\sqrt{1-R}P_{a_3})\rangle<SNL$, $%
\langle\Delta
^2(\sqrt{1-R}P_{a_2}+\sqrt{R}P_{a_3}-P_{a_4})\rangle<SNL$, where SNL
is the shot noise limit. It clearly shows that this state exhibits
the full inseparability (genuine four-party entanglement) according
to the criteria of detecting genuine multipartite CV entanglement
\cite{forteen}. In each of above inequalities, the position and
momentum correlation variance involves three-party combination from
the four entangled modes. Now we compare continuous variables (CV)
four-mode cluster state \cite{fifteen} and GHZ state with the
four-mode entangled state for nonlocal NOPA. The inseparability of
CV four-mode cluster
state are expressed by the correlations of position and momentum of$\langle\Delta ^2(X_1^C+X_2^C+X_3^C)\rangle<SNL$, $%
\langle\Delta ^2(X_3^C+X_4^C)\rangle<SNL$ and $\langle\Delta ^2(P_1^C-P_2^C)\rangle<SNL$, $%
\langle\Delta ^2(P_2^C-P_3^C+P_4^C)\rangle<SNL$
\cite{fifteen,sixteen}, and CV four-partite GHZ state by total
position
$\langle\Delta ^2(X_1+X_2+X_3+X_4)\rangle<SNL$ and relative momentum $%
\langle\Delta ^2(P_i-P_j)\rangle<SNL$ $(i,j=1,2,3,4)$
\cite{forteen,fifteen,sixteen}. We can see that some two-party
correlation of position or momentum components are included in the
criterion inequalities of CV four-partite cluster state and GHZ
state.

%
\begin{figure}
\centerline{
\includegraphics[width=3in]{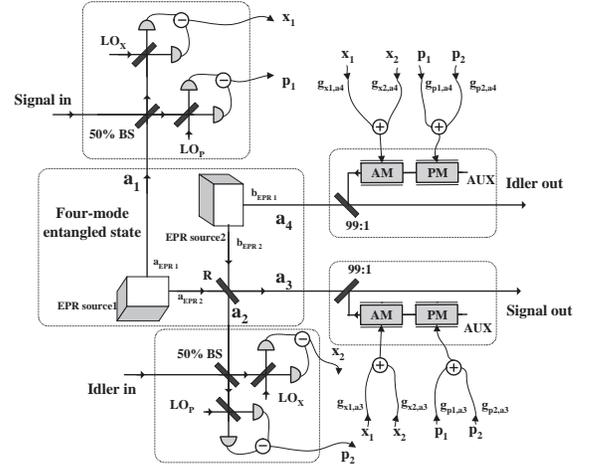}
} \vspace{0.1in}
\caption{ Schematic diagram of nonlocal NOPA with genuine four-mode
entanglement. BS: Beam splitter, LO: Local oscillator, AM: Amplitude
modulator, PM: Phase modulator and AUX: Auxiliary beam. \label{Fig1}
}
\end{figure}

For implementing the nonlocal NOPA the four submodes of the
four-mode entangled state $\hat{a}_{1}, \hat{a}_{2}, \hat{a}_{3},
\hat{a}_{4}$, are distributed into the remote signal input, idler
input, signal output and idler output stations, respectively. The
mode $\hat{a}_{1}$ ($\hat{a}_{2}$) and the input signal
$\hat{a}_{in}^{s}$ (idler $\hat{a}_{in}^{i}$) are combined at a
50/50 beam splitter inside the signal (idler) input station. Then
the two input stations independently perform the homodyne
measurements on themselves two output beams from their 50/50 beam
splitter to extract the amplitude- and phase-quadratures of optical
fields. The obtained quadratures $ (x_1,p_1),(x_2,p_2)$ are
respectively
\begin{eqnarray}
x_1 &=&\frac 1{\sqrt{2}}( \hat{X}_{a_{1}}+\hat{X}_{a_{in}^s}),p_1 =\frac 1{\sqrt{2}}( \hat{P}_{a_{1}}-\hat{P}_{a_{in}^s}), \nonumber \\
x_2 &=&\frac 1{\sqrt{2}}( \hat{X}_{a_{2}}-\hat{X}_{a_{in}^{i}}), p_2
=\frac 1{\sqrt{2}}( \hat{P}_{a_{2}}+\hat{P}_{a_{in}^{i}}). \label{4}
\end{eqnarray}
The input stations send these measured results to both remote output
stations respectively. After receiving these measurement results,
each output station displaces himself mode by means of a 1/99 beam
splitter with an auxiliary beam, the amplitude and phase of which
have been modulated via the amplitude (AM) and phase (PM) modulators
by the sum signals $(x_{a_3},p_{a_3}),(x_{a_4},p_{a_4})$
respectively. The position $(x_{a_3}$ and momentum values $p_{a_3})$
of the modulated sum signals for displacing the entangled mode
$\hat{a}_{3}$ equal to
\begin{eqnarray}
x_{a_3}&=&g_{x1,a_3}x_1+g_{x2,a_3}x_2 \nonumber \\
&=& \sqrt{\frac {1}{1-R}}\hat{X}_{c_{in}^{s}}%
+\sqrt{\frac {R}{1-R}}\hat{X}_{c_{in}^i}-\sqrt{R}\hat{X}_{b_{EPR2}}\nonumber \\
&&+ \sqrt{\frac {1}{1-R}}\hat{X}_{a_{EPR1}}+\frac
{R}{\sqrt{1-R}}\hat{X}_{a_{EPR2}},\nonumber \\
p_{a_3}&=&g_{p1,a_3}p_1+g_{p2,a_3}p_2 \nonumber \\
&=& \sqrt{\frac {1}{1-R}}\hat{P}_{c_{in}^{s}}%
-\sqrt{\frac {R}{1-R}}\hat{P}_{c_{in}^i}-\sqrt{R}\hat{P}_{b_{EPR2}} \nonumber \\
&&-\sqrt{\frac {1}{1-R}}\hat{P}_{a_{EPR1}}+\frac
{R}{\sqrt{1-R}}\hat{P}_{a_{EPR2}},
\end{eqnarray}
where $g_{x1,a_3}=-g_{p1,a_3}=\sqrt{2/(1-R)}$ and $g_{x2,a_3}=
g_{p2,a_3}=-\sqrt{2R/(1-R)}$ are scaling factors, and that for
displacing the entangled beam $\hat{a}_{4}$ equal to
\begin{eqnarray}
x_{a_4}&=&g_{x1,a_4}x_1+g_{x2,a_4}x_2 \nonumber \\
&=& -\sqrt{\frac {R}{1-R}}\hat{X}_{c_{in}^{s}}%
-\sqrt{\frac {1}{1-R}}\hat{X}_{c_{in}^i}+\hat{X}_{b_{EPR2}}\nonumber \\
&&- \sqrt{\frac {R}{1-R}}(\hat{X}_{a_{EPR1}}+\hat{X}_{a_{EPR2}}),\nonumber \\
p_{a_4}&=&g_{p1,a_4}p_1+g_{p2,a_4}p_2 \nonumber \\
&=& \sqrt{\frac {R}{1-R}}\hat{P}_{c_{in}^{s}}%
-\sqrt{\frac {1}{1-R}}\hat{P}_{c_{in}^i}-\hat{P}_{b_{EPR2}} \nonumber \\
&&-\sqrt{\frac {R}{1-R}}(\hat{P}_{a_{EPR1}}-\hat{P}_{a_{EPR2}}),
\end{eqnarray}
where $g_{x1,a_4}=g_{p1,a_4}=-\sqrt{2R/(1-R)}$ and $g_{x2,a_4}=
-g_{p2,a_4}=\sqrt{2/(1-R)}$. Corresponding to the transformation
$\hat{A}\rightarrow \hat{D}^{+}\hat{A}%
\hat{D}=\hat{A}+\left( \hat{X}_m+i\hat{P}_m\right) /2$ in the
Heisenberg representation, the displaced fields at the remote output
stations can be expressed as
\begin{eqnarray}
\hat{a}_{out,a_3}^{s} &=&\sqrt{\frac
{1}{1-R}}\hat{a}_{in}^{s}+\sqrt{\frac
{R}{1-R}}\hat{a}_{in}^{i\dagger} \nonumber \\
&&+\sqrt{\frac
{1}{1-R}}(\hat{a}_{EPR1}^{\dagger}+\hat{a}_{EPR2}),
\nonumber\\
 \hat{a}_{out,a_4}^{i}
&=&\sqrt{\frac {R}{1-R}}\hat{a}_{in}^{s\dagger}+\sqrt{\frac
{1}{1-R}}\hat{a}_{in}^{i}
-(\hat{b}_{EPR2}^{\dagger}+\hat{b}_{EPR1})\nonumber \\
&&+\sqrt{\frac {R}{1-R}}(\hat{a}_{EPR1}+\hat{a}_{EPR2}^{\dagger}).
\label{NPIA}
\end{eqnarray}
Thus we obtain the output signal and idler fields of NOPA
simultaneously (same as Eqs. (\ref{PIA1})and (\ref{PIA2})) at the
remote stations with gain factor $G=1/(1-R)$ in the case of perfect
quantum correlations ($r_1, r_2\rightarrow \infty $). The gain of
NOPA can be adjusted by the reflectivity rate $R$ of the beam
splitter $BS_{0}$. Under the realistic condition with the imperfect
four-mode entanglement, the extra noises will be are added in the
output signals of NOPA.

In conclusion, we have proposed a scheme for nonlocal optical
nondegenerate parametric amplifier using a genuine four-mode
entanglement. This genuine four-mode entanglement is a new class of
multipartite entangled states, which can be generated with two pairs
of EPR entangled beams and linear optics. The inseparability
criteria of the proposed four-mode entangled state are different
from that of CV four-partite cluster state and GHZ state. The
nonlocal NOPA can be implemented by off-line four-mode entanglement,
homodyne detection, classical communication and local displacement.
Due to its nonlocality the proposed NOPA might be extensively
applied in CV quantum information network.

$^{\dagger} $Corresponding author's email address:
jzhang74@yahoo.com, jzhang74@sxu.edu.cn

\section{\textbf{ACKNOWLEDGMENTS}}

This research was supported in part by National Fundamental Research
Natural Science Foundation of China (Grant No. 2006CB921101), NSFC
for Distinguished Young Scholars (Grant No. 10725416), National
Natural Science Foundation of China (Grant No. 60678029, 60736040),
the Cultivation Fund of the Key Scientific and Technical Innovation
Project, Ministry of Education of China (Grant No. 705010), Program
for Changjiang Scholars and Innovative Research Team in University,
Natural Science Foundation of Shanxi Province (Grant No.
2006011003), and the Research Fund for the Returned Abroad Scholars
of Shanxi Province.

\end{document}